\documentclass[doublecol]{epl2} 

\usepackage{graphicx}
\usepackage{dcolumn}
\usepackage{bm}
\usepackage{color}
\usepackage{amsfonts,amssymb,amsmath,latexsym,color,textcomp}
\usepackage{comment}
\usepackage{lineno}

\title{Recurrence Plots 25 years later -- gaining confidence in dynamical transitions}
\shorttitle{Title} 

\author{Norbert Marwan\inst{1,2} \and Stefan Schinkel\inst{3,4} \and J\"urgen Kurths\inst{1,3,5}}
\shortauthor{N. Marwan \etal}

\institute{                    
\inst{1} Potsdam Institute for Climate Impact Research, 14412 Potsdam, Germany\\
\inst{2} Interdisciplinary Center for Dynamics of Complex Systems, University of Potsdam,\\14415 Potsdam, Germany\\
\inst{3} Department of Physics, Humboldt Universit\"at zu Berlin, 10099	Berlin, Germany \\
\inst{4} Department of Psychology, Humboldt Universit\"at zu Berlin, 10099 Berlin, Germany\\
\inst{5} Institute for Complex Systems and Mathematical Biology, University of Aberdeen, UK
}

\date{\today}

\abstract{
Recurrence plot based time series analysis is widely used to study changes and transitions in 
the dynamics of a system or temporal deviations from its overall dynamical regime. However,
most studies do not discuss the significance of the detected variations in the recurrence quantification 
measures. In this letter we propose a novel method to add a confidence measure to the recurrence
quantification analysis. We show how this 
approach can be used to study significant changes in dynamical systems due to a change in 
control parameters, chaos-order as well as chaos-chaos transitions. Finally we study 
and discuss climate transitions by analysing a marine proxy record for past sea surface temperature. \\
This paper is dedicated to the 25th anniversary of the introduction of recurrence plots.
}


\begin{document}

\maketitle

\section{Introduction}
In the November issue of EPL in 1987, Eckmann et al.~proposed the recurrence plot
as a tool to get easily insights into even high-dimensional dynamical systems \cite{eckmann87,marwan2008epjst}. 
Over the last 25 years, their paper has ``led to an active field, with many ramifications [these authors] 
certainly had not anticipated'' \cite{marwan2008epjsteditorial}. Starting from the visual concept
of recurrence plots (RPs), different statistical and quantification approaches have been added, like
recurrence quantification (RQA), dynamical invariants from RPs, and recurrence networks 
\revision{\cite{webber94,marwan2007,marwan2008epjst,donner2010b}}. 25 years after Eckmann's seminal paper, 
RPs and related methods are widely accepted tools for data analysis in various disciplines,
as in physics \cite{vretenar99} and chemistry \cite{rustici99}, but
also for real world systems as in life science \cite{zbilut2004b,marwan2002herz}, 
engineering \cite{nichols2006,montalban2007}, earth science
\cite{marwan2009b}, or finance
and economy \cite{belaire2002,crowley2011,goswami2012}.
This interdisciplinary success is not only caused by the attractive appearance of RPs but
also by the simplicity of the method \cite{webber2009}.
Based on RPs,
we can study the dynamics, transitions, or synchronisation
of complex systems \cite{eckmann87,marwan2007,marwan2008epjst}. In particular, 
such transitions can be uncovered from a changing recurrence 
structure. The different aspects of recurrences can be inferred 
by measures of complexity, also known as {\it recurrence
quantification analysis} (RQA). Although these measures are often
applied on real data and interpreted as indicators of a change
of the system's dynamics, a statistical evaluation of the results
was not yet satisfiably addressed. An early attempt has suggested to use a specific 
model class (e.g., auto-correlated noise) corresponding to the null-hypothesis and 
then testing the RQA results against such models \cite{marwan2003climdyn}.
For a general test of how significant the value of certain RQA measures (in particular
determinism DET and laminarity LAM) is, a test distribution was derived
using binomial distributions \cite{hirata2011a}. 
In order 
to compare time-dependent RQA measures of different observations, a bootstrap
approach was introduced \cite{schinkel2009a}.
However, we still miss a method which can derive the important significance level
of dynamical transitions within {\it one} dynamical system as indicated by RQA. Without providing some 
statement on the confidence of RQA results, any conclusions drawn
from RQA might remain questionable \cite{marwan2011}.

In this letter we propose a method which calculates the confidence level
for the most important, line-based RQA measures. We pick up the idea of
bootstrapping \cite{schinkel2009a} and develop a new algorithm allowing
for gaining confidence in RQA based dynamical transition analysis.
Using this approach we 
for the first time are able to provide a significance statement for
detected transitions of not only qualitatively different systems dynamics based 
on RQA but using only a single observation. This will enable us to interpret the results
of RQA in a more reliable way in the future research and, hence, will further
increase the potentials and acceptance of RQA.

\section{Recurrence Quantification Analysis}

A RP tests for the pair-wise closeness of 
all possible pairs of states $(\vec{x}_i,\vec{x}_j)$ in
an $m$-dimensional phase space, 
$R_{i,j} = \Theta\left(\varepsilon - d(\vec{x}_i,\vec{x}_j)\right)$,
with $\Theta$ as the Heaviside function, $\varepsilon$
as a threshold for closeness \cite{marwan2007,schinkel2008}, and $i, j = 1, \ldots, N$
where $N$ is the number of observed states.
The closeness $d(\vec{x}_i,\vec{x}_j)$ can be measured in 
different ways, using,
e.g., spatial distance, string metric or local rank order \cite{marwan2007}.
Most often, the spatial distance using maximum or Euclidean norm
$d(\vec{x}_i,\vec{x}_j) = \|\vec{x}_i-\vec{x}_j \|$ is used. 
Then, the binary recurrence matrix $\mathbf{R}$ contains the
value one for all close pairs $\|\vec{x}_i-\vec{x}_j \| < \varepsilon$.
A phase space trajectory can be reconstructed from a 
time series by time delay embedding \cite{packard80}.

Similar evolving epochs of the phase space trajectory cause
diagonal structures parallel to the main diagonal in the RP
\cite{marwan2007}. The length
of such diagonal line structures depends on the dynamics of 
the system (periodic, chaotic, stochastic) and can be directly related with dynamically invariant
properties, like $K_2$ entropy \cite{marwan2007}. Therefore, the
distribution $P(l)$ of line lengths $l$ is used by several RQA measures in order to characterise
the system's dynamics \cite{marwan2007}.
Here we focus on the measure {\it determinism} (DET), which is the
fraction of  recurrence points forming diagonal structures,
$DET = \sum_{l=l_{\min}}^N l\, P(l) / \sum_{l=1}^N l\, P(l)$.
A minimal length $l_{\min}$ defines a diagonal 
line \cite{marwan2007}. 
 
Slowly changing states, as occuring during laminar phases 
(intermittency), cause vertical structures in the RP. Therefore,
the distribution $P(v)$ of line lengths $v$ is used to
quantify the laminar phases occuring in a system. Similar to
DET, the measure {\it laminarity} (LAM) is defined as the fraction of the recurrence points 
forming vertical structures,
$LAM = \sum_{v=v_{\min}}^N v\, P(v) / \sum_{v=1}^N v\, P(v)$
\cite{marwan2002herz}. 

The later discussed approach will not only be applicable to these two measures
DET and LAM, but to all line based RQA measures, including recurrence time
based measures \cite{ngamga2012}.


In order to study time dependent behaviour of a system or data
series, we compute these RQA measures using a moving window, applied
on the time series. 
The window has size $w$ and is moved with a step size $s$ 
over the data in such a way that succeeding windows overlap with
$w-s$. This technique was successfully applied to detect
chaos-period transitions \cite{trulla96}, but also more subtle ones
such as chaos-chaos
transitions \cite{marwan2002herz}, or different kinds of
transitions between strange non-chaotic behaviour and
period or chaos \cite{ngamga2007}. It is applicable
to real world data, as demonstrated for the study of, 
e.g., cardiac variability \cite{zbilut2002a}, brain
activity \cite{schinkel2009b}, changes in finance
markets \cite{strozzi2002} or thermodynamic
transitions in corrosion processes \cite{montalban2007}.
However, all these applications miss a clear significance
statement. 

With respect to our goal of a transition detection in the
dynamical system, we formulate the following \emph{null-hypothesis}
$\mathcal{H}_0$:
The dynamics of a system $X$ does not change over time, thus, the recurrence
structure does not change and the RQA measure $M$ of such a system will
therefore be distributed around an unknown, but non-zero mean $\mu(M)$ 
with unknown variance $\sigma(M)$.

For completely random systems the expected distribution of some RQA measures can be modeled \cite{hirata2011a}.
However, for complex real systems it cannot be assumed
that the underlying recurrence structure is completely random but rather
features a certain recurrence structure at all times. 
A dynamical transition in the system changes the recurrence structure 
and, hence, the RQA measures. 
If the impact of the transition is large enough,
it will push the RQA measure $M$ out of its normal range. 
The deviation from this normal range  can be considered as significant if
the observed value of $M(t)$ at time point $t$ is outside
of a predefined interquantile range such as $[\alpha/2,\ 1-\alpha /2]$.

\section{Variance estimation by bootstrapping}
\noindent 
In order to test for significant deviations from 
the unknown mean of the data, we first have to estimate
the variance of the RQA measures in question. To do so,
we introduce a bootstrap approach in the calculation of the RQA
measures \cite{efron1998}. 
Bootstrapping is a conceptually simple yet powerful
statistical tool to estimate the variance of statistical parameters, such 
as the mean, even if the underlying distribution 
is unknown. Since we cannot assume
that the distributions of line lengths $P(l)$ and $P(w)$
follow a known probability distribution, we use
this advantage of the bootstrap approach to estimate the confidence 
bounds of the RQA measures $M$ which rely on these distributions.
We will use bootstrap resampling to create a test distribution
of the RQA measures from which we can then estimate the overall
mean and variance of those measures and, finally, to formulate
the important significance statement.


The time dependent RQA analysis is based on moving
windows, shifted over the time series, and calculating the
RP within these windows.
For each of the $N_{\rm w}$ time steps of the moving
window $t$ ($t = w/2, 3w/2, 5w/2, \dots, N-w/2$), 
i.e., for different time points,
we get the local RPs with $n(t)$ diagonal lines
and then calculate the corresponding local histograms of diagonal lines $P_t(l)$.
The time dependent RQA measures $M(t)$ (e.g., $M(t) =  DET(t)$ are calculated from  $P_t(l)$.

In order to estimate a general distribution of the RQA measures
following our null-hypothesis $\mathcal{H}_0$, we suggest the following 
procedure. All local histograms $P_t(l)$ are merged together 
in order to get an overall histogram and thus a statistical
average of the recurrence structure of the system, i.e., 
we bootstrap from the unification 
\begin{equation}
\hat P(l) = \sum_t P_t(l)
\end{equation}
of the local histograms.
We draw $\bar n$ recurrence structures (i.e.~diagonal lines) 
from $\hat P(l)$. The number $\bar n$ of 
drawings is the mean number of recurrence structures $n(t)$
contained in the local distributions $P_t(l)$, 
\begin{equation}
\bar n = \frac{1}{N_{\rm w}} \sum_{t=w/2}^{N-w/2}  n(t) = \frac{1}{N_{\rm w}} \sum_{t=w/2}^{N-w/2} \sum_{l=1}^N P_t(l).
\end{equation}
From the resulting empirical distribution $P(l)^*$,
we compute the corresponding RQA measure, say in our case DET.
By repeating this procedure $B$ times (e.g.~$B=1,000$), we
get the test distribution for DET, say $F(DET)$. By 
calculation of the $\alpha$-quantiles of the
distribution $F(DET)$, we derive the confidence intervals
of DET which can be used to statistically infer the
significance of the changes of $DET(t)$, and thus the
observed transitions.

\section{Illustration of the method}

\noindent We illustrate the proposed statistical
test on two model systems: (1) a linear autocorrelated 
process and (2) a nonlinear process, both for changing
parameters.

\begin{figure}[h] 
  \centering
\includegraphics[width=\columnwidth]{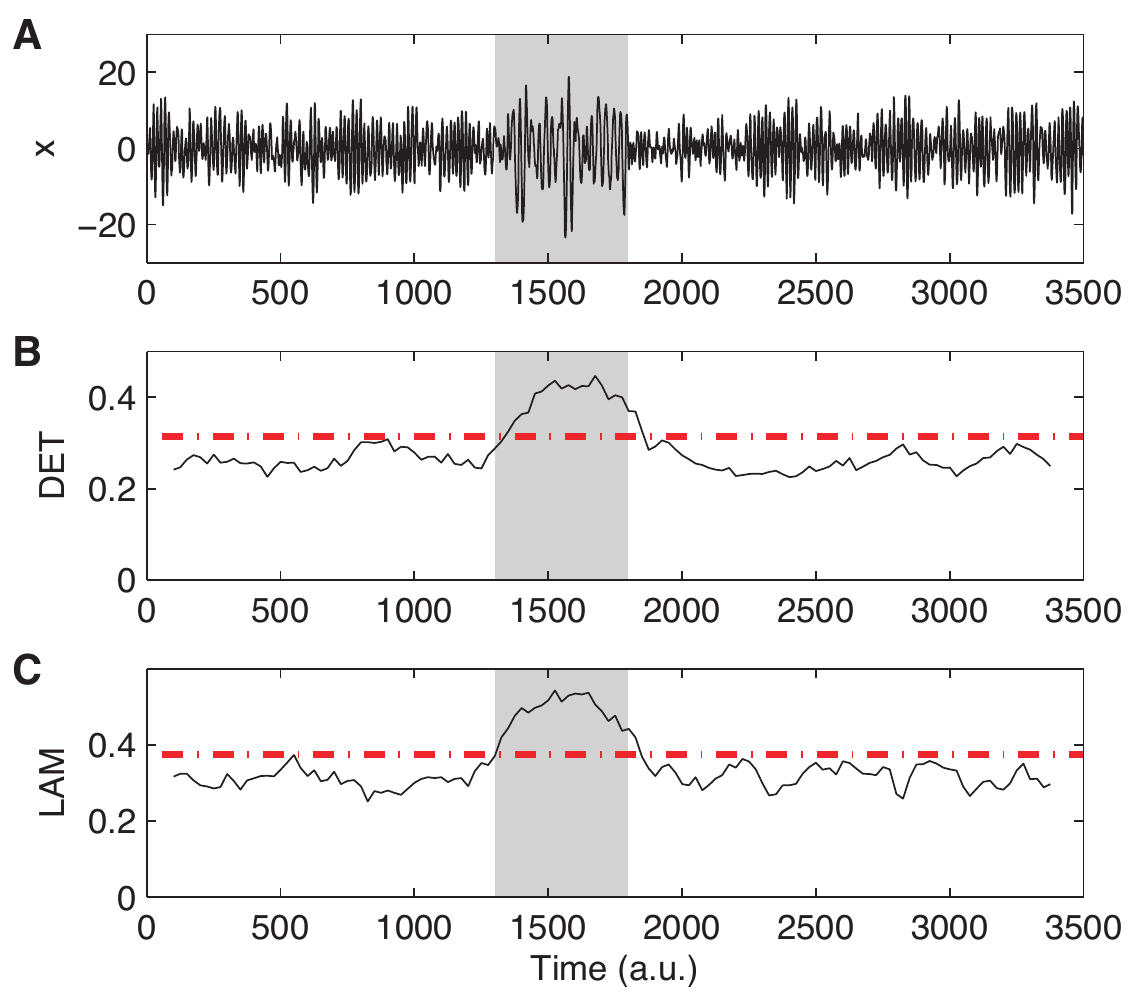} 
\caption{(Colour online) (A) Autocorrelated process with slight transitions 
of the parameters between time 1,300 and 1,800 (shaded region).
Corresponding RQA measures (B) determinism DET
and (C) laminarity LAM, indicating the epoch of changed 
dynamics in the autocorrelated process between 1,300 and 1,800
(shaded area) by an increasing of their values. This increase exceeds the
99\% confidence interval (dashed line) as derived by the proposed
bootstrapping approach.
}\label{fig_rqa_AR}
\end{figure}

(1) Our first example is an autocorrelated stochastic signal with changing
properties, i.e., an autoregressive
process of order 2 
\begin{equation}
x_{i} = a_1  x(i-1) + a_2  x(i-2) + b \xi(i)
\end{equation}
with $a_1 = 1.80$, $a_2 = -0.972$ and $b = 0.64$.
After time step 1,300, the AR coefficients slightly change
to $a_1 = 1.85$, $a_2 = -0.917$ and $b = 0.76$ for 500
time steps. Afterwards these coefficients are changed
back to the initial values. With this procedure
the signal contains a short epoch of slightly changed dynamics
(Fig.~\ref{fig_rqa_AR}A).

Next we compute the RQA measures DET and LAM
from this data series (no embedding) using windows
of size $w=200$ and with a step size of $s=25$. The 
threshold $\varepsilon$ is chosen for each window
separately to preserve a constant recurrence
rate of 7.5\% \cite{schinkel2008}. 
The bootstrap resampling is then applied using 1,000
resamplings. As we expect in the window of increased 
auto-correlation a larger number of diagonal and vertical
lines, we will only consider the upper confidence level.

The DET measure reveals a high number of diagonal lines 
in the RP. Before time 1,300 and after time 1,800,
DET values vary between 0.25 and 0.3. This coincides
with the moderate auto-correlation of the process.
Between the time 1,300 and 1,800, DET shows an
increase and exceeds the confidence interval of
0.31, corresponding to a 99\% confidence level. Similar,
LAM varies before and after the inset of changed 
dynamics at a lower level ($LAM \approx 0.3$) and
increases within the period between time 1,300 and 1,800  up
to $LAM \approx 0.55$ due to its increased persistence. This
increase of DET and LAM confirms the further increase of the
auto-correlation of the considered process within this
epoch.

\begin{figure}[h!] 
  \centering
\includegraphics[width=\columnwidth]{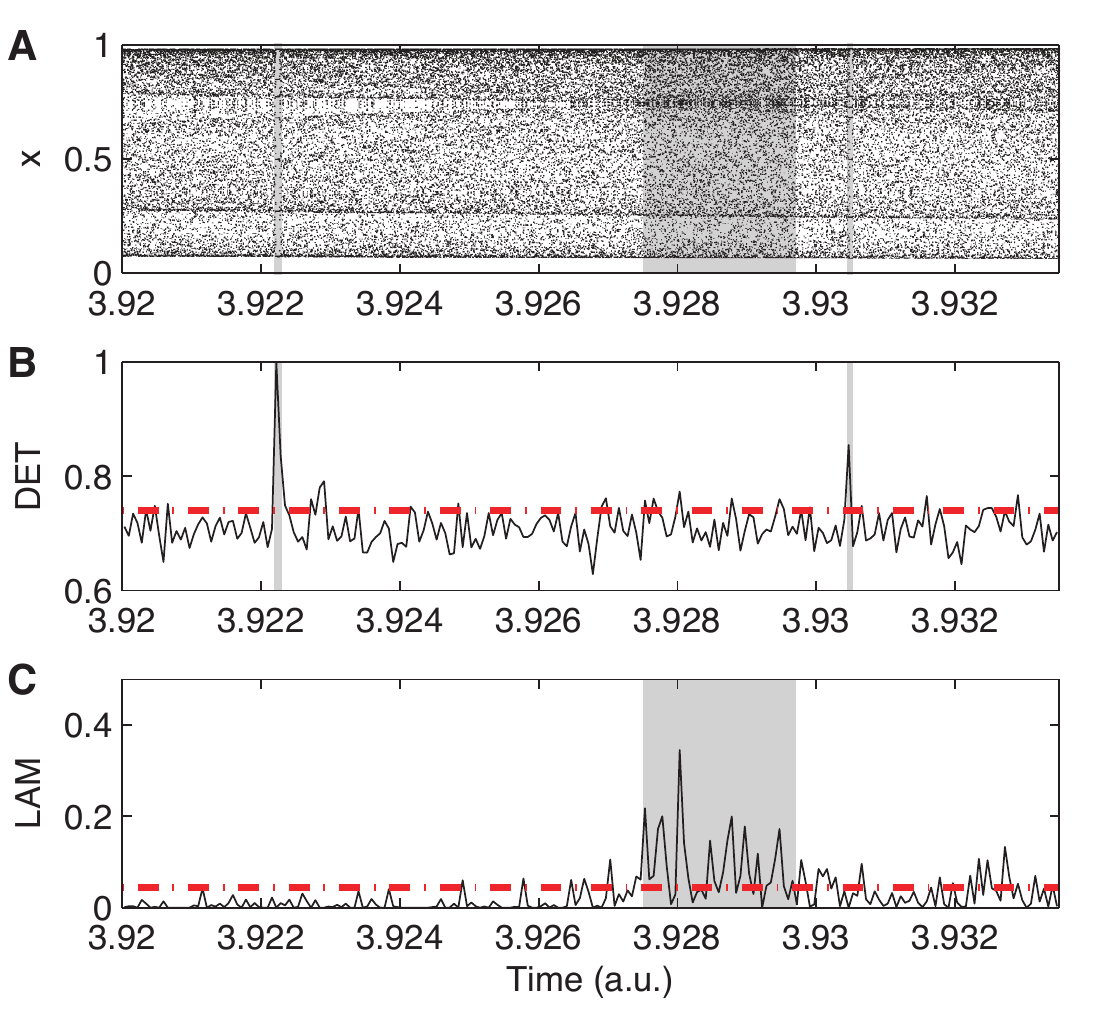}
\caption{(A) Logistic map with chaos-period and chaos-chaos
transitions for control parameter $a = [3.9200\ 3.9335]$
and corresponding RQA measures (B) DET and (C) LAM. 
For $a=[3.92221\ 3.92227]$ we have 
a period-7 window, for $a=[3.93047\ 3.93050]$
a period-8 window and at a broad range around
$a=3.928$ intermittency (marked with shaded area).
(B) and (C) 99\% significance levels are shown as dash-dotted lines.
}\label{fig_log}
\end{figure}

(2) To test whether the proposed method is also capable
of providing a quanatitative statement of more subtle
changes in dynamics, like chaos-order and chaos-chaos transitions, 
we use a modified logistic map
with mutual transitions \cite{trulla96}
\begin{equation}
x_{i+1} = a(i)  x(i)  (1 - x(i))
\end{equation}
with the control parameter $a$ in the range
$[3.9200\ 3.9335]$ with increments of $\Delta a = 2.5\ 10^{-7}$.
Using this intervall we find for $a=[3.92221\ 3.92227]$
a period-7 window, for $a=[3.93047\ 3.93050]$
a period-8 window and at a broad range around
$a=3.928\dots$ intermittency
(Fig.~\ref{fig_log}). Again, for these kind of dynamical transitions
we can expect increased values of DET and LAM, hence,
we only need to consider the upper confidence level.

Next we compute the RQA measures DET and LAM 
from this data series (no embedding) using windows
of size $w=250$ and with a step size of $s=250$. The 
threshold $\varepsilon$ is chosen for each window
separately in order to preserve a constant recurrence
rate of 5\%. As a line structure we consider each line
with a length of at least two points, 
i.e.~$l_{\min} = v_{\min} = 2$.

The measure DET shows for the periodic windows at
$a=[3.92221\ 3.92227]$ and $a=[3.93047\ 3.93050]$
maxima \cite{marwan2002herz}. The periodic behaviour of the system causes
only long diagonal lines, resulting in high values of DET.
In contrast, LAM shows high values only for the 
region of intermittency around $a=3.928\dots$. In this
region, the system has slowly changing, laminar states \cite{marwan2002herz}.
For the proposed bootstrapping approach, we use 1,000 resamplings
in order to construct the test statistics. 
As the 99\%-quantile we find
for DET $q_{0.99} = 0.74$ and for LAM $q_{0.99} = 0.04$.
These values provide the 99\% confidence level for DET and LAM.
Thus, the two maxima of DET in the periodic windows are 
significant on a 99\% level ( p $<$ 0.01). For LAM we find several significant
high values of 99\% significance in the region of intermittency
around $a=3.928$. This is due to the longer
range of intermittent behaviour in this region of the
control parameter $a$.

%
%

\section{Application to real world data}

\begin{figure}[th] 
  \centering
\onefigure[width=\columnwidth]{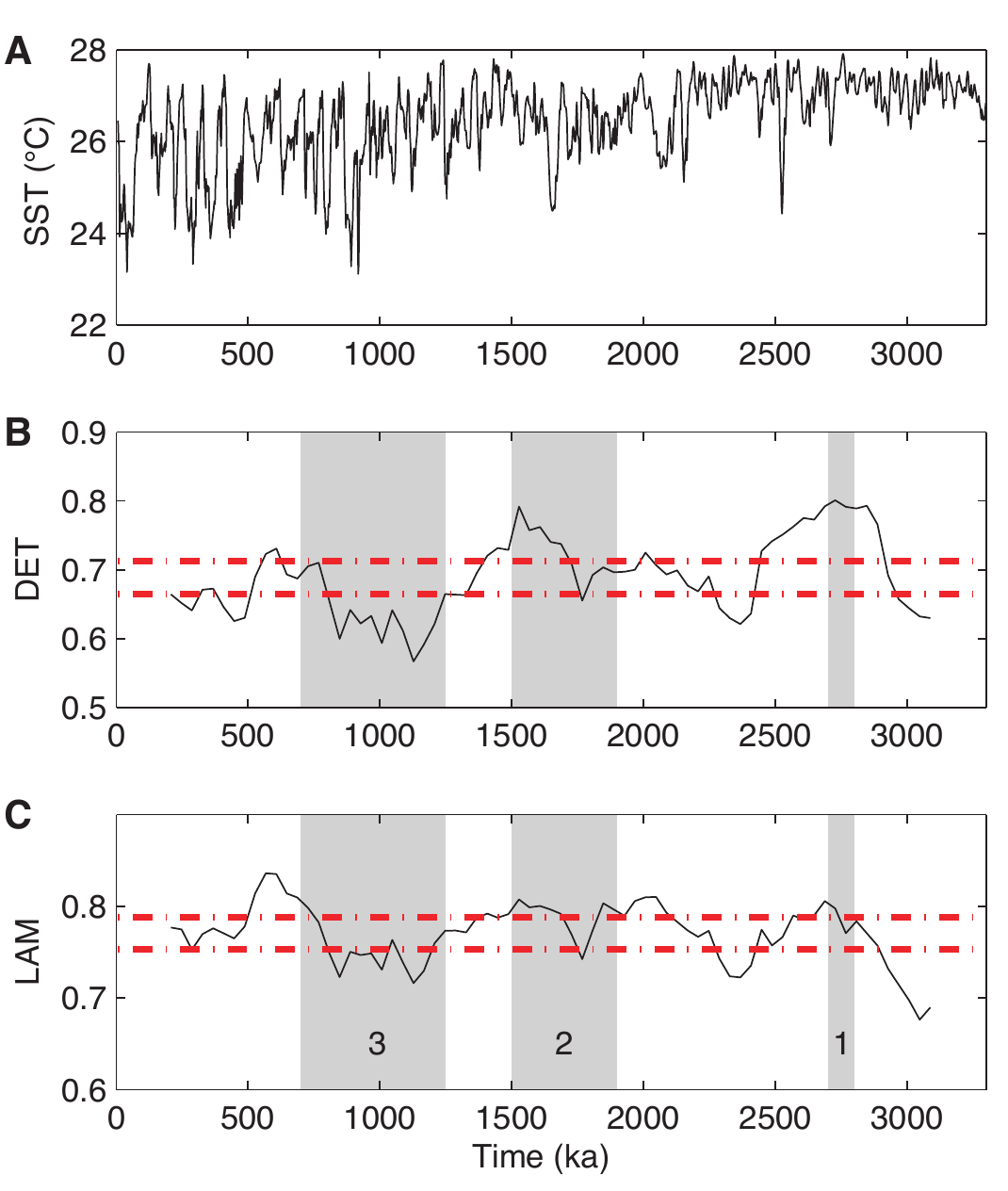}
\caption{(A) Alkone SST record of ODP site 722, 
and corresponding (B) DET and (C) LAM 
measures (95\% confidence bounds are shown with blue
dash-dotted lines).
The transition to (1) the Northern Hemisphere glaciation,
the (2) intensification of the Walker circulation, and (3) the
transition phase from glaciation to glacial-interglacial cycles
are marked by shaded areas.
}\label{fig_rqa_odp722}
\end{figure}

The climate system is a highly complex one which has undergone various
transitions in the past.
The investigation of relationships between sea surface temperature (SST) and specific
climate responses, like the Asian monsoon system or the thermohaline circulation
in the Atlantic, represents an important scientific challenge for
understanding the global climate system, its mechanisms, and its related variability. 
In palaeoclimatology, different archives are used to reconstruct and study climate
conditions of the past, as lake \cite{marwan2003climdyn} and marine sediments \cite{herbert2010} or
speleothemes \cite{kennett2012}. Alkenone remnants in the organic fraction of
marine sediments, produced
by phytoplankton, can be used to reconstruct SST of the past, allowing to study 
the temperature variability of the oceans \cite{herbert2001}.
Here we will use a marine record from the Ocean Drilling Programme
(ODP) derived from a drilling in the Arabian see, ODP site 722. This record provides
alkenone based reconstructed SST in the realm of the Asian monsoon system
for the past 3.3~Ma (Fig.~\ref{fig_rqa_odp722}A) \cite{herbert2010}.
During this epoch, a dramatic climate change happened by two steps of
global cooling \cite{herbert2010}. The first step between 3.0 and 2.5~Ma 
coincides with the high-latitude Northern Hemisphere glaciation. 
The second step of cooling occurred between 2.0 and 1.5~Ma
and is related with a continuous cooling of the subtropical oceans but a stationary
high-latitude climate.
Some mechanisms of these global-scale climate changes are known
and coincide with a transition to an obliquity-driven climate variability
with a 41~ka period after 2.8--2.7~Ma \cite{herbert2010}, 
a shift from that climate variability (with high-latitude glaciation) 
to glacial-interglacial cycles with a 100~ka period after
a transition period between 1.25 and 0.7~Ma \cite{mudelsee1997}, 
and the development of the Walker circulation at 1.9--1.5~Ma \cite{ravelo2004}.
The RQA and the proposed significance
test are promising tools to analyze the alkenone SST record of the ODP site 722.

The original time series of ODP 722 is not equally sampled. Therefore, we interpolate
it to a time series with sampling period of 2~ka. For performing the RQA
we  use a time delay embedding with dimension $m=3$ and delay 
$\tau=2$. The threshold is chosen to preserve a constant recurrence
rate of 7.5\%. The bootstrapping is performed using 1,000 resamplings.
In this real world example, we use a reduced confidence level
of 95\%. As we do not know which kinds of dynamical transition are there,
we will consider both the upper and the lower confidence level.

The RQA measures DET and LAM reveal various significantly high and low values
as summarised in Fig.~\ref{fig_rqa_odp722} and Tab.~\ref{tab_rqa_odp722}.

Around 3.0~Ma ago, a long-lasting period of warm climate
with a permanent El Ni\~no came to an end. This general change from a 
warm climate towards a more variable and cooler one
is clearly indicated by a change from low to high DET and LAM
values.
 
The first cooling phase between 3.0 and 2.5~Ma is well indicated by 
high values of the measure DET which can be considered to reflect an
increase in regularity and auto-correlation of the system.
The rapid onset of the Northern Hemisphere glaciation between 2.8 and 2.7~Ma
is marked by an increase of LAM, corresponding to an intermittent behaviour. 
The fact that DET and, thus, the auto-correlation increased before the 
intermittent behaviour can be
understood as a critical slowing down of the dynamics as it is typical
for tipping points \cite{dakos2008}. The increase of DET
might, therefore, be an indication that the climate system reached a
tipping point at 3.0--2.9~Ma, leading to the regime change of Northern Hemisphere 
glaciation.

Between 2.4 and 2.3~Ma, DET and LAM decreased, revealing a short
period of more irregular and stochastic variability. This might be an 
indication for a transition between two different regimes. This transition
was not yet found in palaeoclimate literature, but is confirmed by
another study also using a nonlinear measure for transition
detection \cite{malik2012b}.

The period of the development of the Walker circulation between 
1.9 and 1.5~Ma is marked by an increase in both, DET and LAM.

The transition period from the glaciation regime with dominant 
41~ka cycle to the glacial-interglacial regime with 100~ka cycle 
is marked by a significant decrease of the measures DET and LAM.
This corresponds to a phase of less regularity or more stochastic
variability of the SST.

Further high values in DET and LAM occur at around 2.0~Ma and
between 0.75 and 0.5~Ma. At 2.0~Ma, a reorganization of
subtropical and tropical ocean circulation begun which was triggered
by high-latitude cooling and its impact on deepwater formation.
Between 0.75 and 0.5~Ma, the sensitivity of the high-latitude
climate response to solar forcing reached its maximum \cite{ravelo2004}.
This is consistent with recent findings of a coherence between solar forcing
and climate variability in this region during this period \cite{malik2012b}.

The transitions found correspond to dynamical transitions caused 
by different changes in climate. The recurrence based analysis can not only detect
these transitions but also provide additional information about the climate transitions,
whose onsets are, at least partly, known \cite{donges2011d,malik2012b}.


\begin{table*}[htdp]
\caption{Major regime changes in alkenone SST record from ODP site 722 as indicated
by significant high values of determinism (DET$+$) and laminarity (LAM$+$) as
well as significant low values (DET$-$ and LAM$-$).}\label{tab_rqa_odp722}
\begin{center}
\begin{tabular}{lrrrr}
Period	&DET$+$	&DET$-$	&LAM$+$		&LAM$-$\\
\hline
\hline
Northern Hemisphere glaciation	&2.9--2.5	&		&2.75--2.65	&\\	
\hline
Interregime transition		&		&2.4--2.3	&			&2.4--2.3\\
\hline
(Sub-)Tropical reorganisation	&2.0		&		&2.05--1.95	&\\
\hline
Development Walker circulation	&1.7--1.4	&		&1.7--1.5		&\\
\hline
Transition 41~ka to 100~ka	&		&1.25--0.8&			&1.2--0.8\\	
\hline
Maximal climate sensitivity	&0.65-0.55&		&0.75--0.5	&\\
\hline
\end{tabular}
\end{center}
\label{default}
\end{table*}%

\section{Conclusion}
\noindent We have introduced a bootstrap based approach for providing confidence levels for line-based
recurrence quantification measures, which are related to dynamical properties (like Lyapunov exponent or 
$K_2$ entropy). Using this technique, we are able to investigate changing dynamics
by RQA and can, for the first time, provide confidence levels for the variation of the RQA measures
and, thus, the changed dynamics. 
We have shown the potential of the approach by studying dynamical changes in an
auto-correlated process and for chaos-order and chaos-chaos transitions.
These examples have also demonstrated
the importance of considering confidence intervals, as fluctuations in the RQA measures 
can be misinterpreted if the overall variance of these measures is not taken into consideration.

The application of our approach on sea surface temperature variability of the past has demonstrated
that recurrence based analysis provides new insights in known palaeo-climate changes. Recurrence 
properties can be help for a better understanding of the mechanisms of the transitions between
different climate regimes.

25 years after the introduction of recurrence plots by Eckmann et al.~\cite{eckmann87}, the development
of this technique still continues. With our paper we would like to honor the seminal work by these
authors, but would also like to emphasize that the calculation of confidence levels for the RQA
measures is an important requirement for the method to get widely accepted. It is highly desirable
that future research using RQA comes along with corresponding confidence levels.

\acknowledgments
\noindent This work was supported by the Potsdam Research Cluster for Georisk Analysis,
Environmental Change and Sustainability  
(PROGRESS, BMBF support code 03IS2191B), the DFG research groups FOR 1380 (HIMPAC) and 
FOR 868 (``Computational Modeling of Behavioral, Cognitive, and Neural Dynamics'').


\begin{thebibliography}{10}
\expandafter\ifx\csname url\endcsname\relax\def\url#1{\texttt{#1}}\fi

\bibitem{eckmann87}
\Name{Eckmann J.-P., {Oliffson Kamphorst} S. \and Ruelle D.}
  \REVIEW{Europhysics Letters}{5}{1987}{973}.

\bibitem{marwan2008epjst}
\Name{Marwan N.} \REVIEW{European Physical Journal -- Special
  Topics}{164}{2008}{3}.

\bibitem{marwan2008epjsteditorial}
\Name{Marwan N., Facchini A., Thiel M., Zbilut J.~P. \and Kantz H.}
  \REVIEW{European Physical Journal -- Special Topics}{164}{2008}{1}.

\bibitem{webber94}
\Name{{Webber Jr.} C.~L. \and Zbilut J.~P.} \REVIEW{Journal of Applied
  Physiology}{76}{1994}{965}.
\Name{Zbilut J.~P. \and {Webber Jr.} C.~L.} \REVIEW{International Journal of
  Bifurcation and Chaos}{17}{2007}{3477}.

\bibitem{marwan2007}
\Name{Marwan N., Romano M.~C., Thiel M. \and Kurths J.} \REVIEW{Physics
  Reports}{438}{2007}{237}.

\bibitem{donner2010b}
\Name{Donner R.~V., Zou Y., Donges J.~F., Marwan N. \and Kurths J.} \REVIEW{New
  Journal of Physics}{12}{2010}{033025}.

\bibitem{vretenar99}
\Name{Vretenar D., Paar N., Ring P. \and Lalazissis G.~A.} \REVIEW{Physical
  Review E}{60}{1999}{308}.
\Name{Jamitzky F., Stark M., Bunk W., Heckl W.~M. \and Stark R.~W.}
  \REVIEW{Nanotechnology}{17}{2006}{S213}.

\bibitem{rustici99}
\Name{Rustici M., Caravati C., Petretto E., Branca M. \and Marchettini N.}
  \REVIEW{Journal of Physical Chemistry A}{103}{1999}{6564}.
\Name{{Garc\'ia-Ochoa} E., {Gonz\'alez-S\'anchez} J., na N.~A. \and Euan J.}
  \REVIEW{Journal of Applied Electrochemistry}{39}{2009}{637}.

\bibitem{marwan2002herz}
\Name{Marwan N., Wessel N., Meyerfeldt U., Schirdewan A. \and Kurths J.}
  \REVIEW{{Physical Review E}}{66}{2002}{026702}.

\bibitem{zbilut2004b}
\Name{Zbilut J.~P., Giuliani A., Colosimo A., Mitchell J.~C., Colafranceschi
  M., Marwan N., Uversky V.~N. \and {Webber Jr.} C.~L.} \REVIEW{Journal of
  Proteome Research}{3}{2004}{1243}.
\Name{Stam C.~J.} \REVIEW{Clinical Neurophysiology}{116}{2005}{2266}.
\Name{Schinkel S., Marwan N. \and Kurths J.} \REVIEW{Cognitive
  Neurodynamics}{1}{2007}{317}.

\bibitem{nichols2006}
\Name{Nichols J.~M., Trickey S.~T. \and Seaver M.} \REVIEW{Mechanical Systems
  and Signal Processing}{20}{2006}{421}.
\Name{Sen A.~K., Longwic R., Litak G. \and G\'orski K.} \REVIEW{Mechanical
  Systems and Signal Processing}{22}{2008}{362}.
  
\bibitem{montalban2007}
\Name{Montalb\'an L.~S., Henttu P. \and Pich\'e R.} \REVIEW{International
  Journal of Bifurcation and Chaos}{17}{2007}{3725}.

\bibitem{marwan2009b}
\Name{Marwan N., Donges J.~F., Zou Y., Donner R.~V. \and Kurths J.}
  \REVIEW{Physics Letters A}{373}{2009}{4246}.
\Name{March T.~K., Chapman S.~C. \and Dendy R.~O.} \REVIEW{Geophysical Research
  Letters}{32}{2005}{1}.
\Name{Zolotova N.~V. \and Ponyavin D.~I.} \REVIEW{Solar
  Physics}{243}{2007}{193}.

\bibitem{belaire2002}
\Name{Belaire-Franch J., Contreras D. \and {Tordera-Lled\'o} L.}
  \REVIEW{Physica D}{171}{2002}{249}.

\bibitem{crowley2011}
\Name{Crowley P.~M. \and Schultz A.} \REVIEW{International Journal of
  Bifurcation and Chaos}{21}{2011}{1215}.

\bibitem{goswami2012}
\Name{Goswami B., Ambika G., Marwan N. \and Kurths J.} \REVIEW{Physica
  A}{391}{2012}{4364}.

\bibitem{webber2009}
\Name{{Webber Jr.} C.~L., Marwan N., Facchini A. \and Giuliani A.}
  \REVIEW{Physics Letters A}{373}{2009}{3753}.

\bibitem{marwan2003climdyn}
\Name{Marwan N., Trauth M.~H., Vuille M. \and Kurths J.} \REVIEW{{Climate
  Dynamics}}{21}{2003}{317}.

\bibitem{hirata2011a}
\Name{Hirata Y. \and Aihara K.} \REVIEW{International Journal of Bifurcation
  and Chaos}{21}{2011}{1077}.

\bibitem{schinkel2009a}
\Name{Schinkel S., Marwan N., Dimigen O. \and Kurths J.} \REVIEW{Physics
  Letters A}{373}{2009}{2245}.

\bibitem{marwan2011}
\Name{Marwan N.} \REVIEW{International Journal of Bifurcation and
  Chaos}{21}{2011}{1003}.

\bibitem{schinkel2008}
\Name{Schinkel S., Dimigen O. \and Marwan N.} \REVIEW{European Physical Journal
  -- Special Topics}{164}{2008}{45}.

\bibitem{packard80}
\Name{Packard N.~H., Crutchfield J.~P., Farmer J.~D. \and Shaw R.~S.}
  \REVIEW{Physical Review Letters}{45}{1980}{712}.

\bibitem{ngamga2012}
\Name{Ngamga E.~J., Senthilkumar D.~V., Prasad A., Parmananda P., Marwan N.
  \and Kurths J.} \REVIEW{Physical Review E}{85}{2012}{026217}.

\bibitem{trulla96}
\Name{Trulla L.~L., Giuliani A., Zbilut J.~P. \and {Webber Jr.} C.~L.}
  \REVIEW{Physics Letters A}{223}{1996}{255}.

\bibitem{ngamga2007}
\Name{Ngamga E.~J., Nandi A., Ramaswamy R., Romano M.~C., Thiel M. \and Kurths
  J.} \REVIEW{Physical Review E}{75}{2007}{036222}.

\bibitem{zbilut2002a}
\Name{Zbilut J.~P., Thomasson N. \and {Webber Jr.} C.~L.} \REVIEW{Medical
  Engineering \& Physics}{24}{2002}{53}.

\bibitem{schinkel2009b}
\Name{Schinkel S., Marwan N. \and Kurths J.} \REVIEW{Journal of
  Physiology-Paris}{103}{2009}{315}.

\bibitem{strozzi2002}
\Name{Strozzi F., {Zald\'ivar} J.-M. \and Zbilut J.~P.} \REVIEW{Physica
  A}{312}{2002}{520}.

\bibitem{efron1998}
\Name{Efron B. \and Tibshirani R.~J.} \Book{{An Introduction to the Bootstrap}}
  (Chapman \& Hall/CRC, Boca Raton, London, New York, Washington DC) 1998.

\bibitem{herbert2010}
\Name{Herbert T.~D., Peterson L.~C., Lawrence K.~T. \and Liu Z.}
  \REVIEW{Science (New York, N.Y.)}{328}{2010}{1530}.

\bibitem{kennett2012}
\Name{Kennett D.~J., Breitenbach S. F.~M., Aquino V.~V., Asmerom Y., Awe J.,
  Baldini J. U.~L., Bartlein P., Culleton B.~J., Ebert C., Jazwa C., Macri
  M.~J., Marwan N., Polyak V., Prufer K.~M., Ridley H.~E., Sodemann H.,
  Winterhalder B. \and Haug G.~H.} \REVIEW{Science}{338}{2012}{788}.

\bibitem{herbert2001}
\Name{Herbert T.~D.} \REVIEW{Geochemistry Geophysics
  Geosystems}{2}{2001}{1005}.

\bibitem{mudelsee1997}
\Name{Mudelsee M. \and Schulz M.} \REVIEW{Earth and Planetary Science
  Letters}{151}{1997}{117}.

\bibitem{ravelo2004}
\Name{Ravelo A.~C., Andreasen D.~H., Lyle M., Lyle A.~O. \and Wara M.~W.}
  \REVIEW{Nature}{429}{2004}{263}.

\bibitem{dakos2008}
\Name{Dakos V., Scheffer M., van Nes E.~H., Brovkin V., Petoukhov V. \and Held
  H.} \REVIEW{Proceedings of the National Academy of Sciences of the United
  States of America}{105}{2008}{14308}.
\Name{Scheffer M., Bascompte J., Brock W.~A., Brovkin V., Carpenter S.~R.,
  Dakos V., Held H., van Nes E.~H., Rietkerk M. \and Sugihara G.}
  \REVIEW{Nature}{461}{2009}{53}.

\bibitem{malik2012b}
\Name{Malik N., Zou Y., Marwan N. \and Kurths J.} \REVIEW{Europhysics Letters
  (EPL)}{97}{2012}{40009}.

\bibitem{donges2011d}
\Name{Donges J.~F., Donner R.~V., Trauth M.~H., Marwan N., Schellnhuber H.~J.
  \and Kurths J.} \REVIEW{Proceedings of the National Academy of
  Sciences}{108}{2011}{20422}.

\end{thebibliography}
\end{document}